\documentclass[acmsmall,screen]{acmart}

\acmJournal{PACMPL}
\acmVolume{1}
\acmNumber{CONF} 
\acmArticle{1}
\acmYear{2018}
\acmMonth{1}
\acmDOI{} 
\startPage{1}

\setcopyright{none}

\usepackage{graphicx}        
\usepackage{algorithm}       
\usepackage[noend]{algpseudocode}   
\usepackage{cleveref}        
\usepackage{xcolor}          
\usepackage{caption}
\usepackage{subcaption}
\usepackage{microtype}
\usepackage{semantic}
\usepackage{multirow}
\usepackage{siunitx}
\usepackage{setspace}
\usepackage{xspace}
\usepackage{enumitem}

\newcommand\todo[1]{{\color{red}{#1}}}

\newcommand{\bblb}{\textnormal{[\kern-.15em[}}
\newcommand{\bbrb}{\textnormal{]\kern-.15em]}}

\renewcommand{\eval}[1]{\ensuremath{\bblb#1\bbrb}}
\newcommand{\aeval}[1]{\ensuremath{\bblb#1\bbrb}^{\sharp}}

\newcommand{\name}{\textsc{SyMetric}\xspace}
\DeclareMathOperator*{\argmin}{arg\,min}

\begin{document}

\title{Metric Program Synthesis}
\author{John Feser}
\orcid{0000-0001-8577-1784}
\affiliation{
	\institution{MIT}
	\department{CSAIL}
	\streetaddress{32 Vassar St}
	\city{Cambridge}
	\country{USA}
}
\email{feser@mit.edu}

\author{Isil Dillig}
\affiliation{
	\institution{UT Austin}
	\city{Austin}
	\country{USA}
}
\email{isil@cs.utexas.edu}

\author{Armando Solar-Lezama}
\affiliation{
	\institution{MIT}
	\department{CSAIL}
	\streetaddress{32 Vassar St}
	\city{Cambridge}
	\country{USA}
}
\email{asolar@mit.edu}

\begin{abstract}
	We present a new domain-agnostic synthesis technique for generating programs from input-output examples.
	Our method, called \emph{metric program synthesis}, relaxes the well-known \emph{observational equivalence} idea (used widely in bottom-up enumerative synthesis) into a weaker notion of \emph{observational similarity}, with the goal of reducing the search space that the synthesizer needs to explore. Our method clusters  programs into equivalence classes based on a \emph{distance metric} and constructs a version space that compactly represents ``approximately correct'' programs. Then, given a ``close enough'' program sampled from this version space, our approach uses a  distance-guided repair algorithm to find a program that exactly matches the given input-output examples.	We have implemented our proposed metric program synthesis technique in a tool called \name and evaluate it in three different domains considered in prior work.  Our evaluation shows that \name{} outperforms other domain-agnostic synthesizers that use observational equivalence and that it achieves results  competitive with domain-specific synthesizers that are either designed for or trained on those domains.

\end{abstract}

\maketitle
\newcommand{\fta}{\ensuremath{\mathcal{A}}}
\newcommand{\ftastate}{\ensuremath{q}}
\newcommand{\ftastates}{\ensuremath{Q}}
\newcommand{\alphabet}{\ensuremath{\Sigma}}
\newcommand{\wildcard}{\ensuremath{\bot}}
\newcommand{\nodes}{\ensuremath{N}}
\newcommand{\rootnode}{\ensuremath{\mathsf{root}}\xspace}
\newcommand{\finalstates}{\ensuremath{\ftastates_f}}
\newcommand{\transitions}{\ensuremath{\Delta}}
\newcommand{\transition}{\ensuremath{\delta}}
\newcommand{\Subs}{\ensuremath{\mathsf{SubTrees}}}
\newcommand{\LanguageOf}[1]{\ensuremath{\mathcal{L}(#1)}}
\newcommand{\true}{\emph{true}}
\newcommand{\false}{\emph{false}}

\section{Introduction}\label{sec:intro}

\emph{Programming-by-example (PBE)} is a program synthesis task wherein the goal is to learn a program in some domain-specific language (DSL) that is consistent with a given set of input-output examples.  Because of its potential to democratize programming for computer end-users, PBE has attracted enormous attention from several research communities and has found a number of useful applications ranging from string and table transformations~\cite{flashfill,morpheus} to question answering~\cite{web-qa} to computer-aided design (CAD)~\cite{Du2018}.

While there are many different techniques to solve the PBE problem, existing solutions can be classified as either being \emph{domain-agnostic} or \emph{domain-specific}. Domain-specific techniques (e.g.,~\cite{flashfill,Du2018,morpheus,web-qa,regel,scythe,lambda2}) are specialized to a particular DSL and target a pre-defined class of synthesis tasks. Domain-agnostic techniques~\cite{sketch,neo,blaze} are parameterized over a DSL and can, in principle, be applied to a wide variety of  domains.

A common domain-agnostic solution to the PBE problem is to perform \emph{bottom-up enumeration} over DSL programs~\cite{transit,blaze,burst}. The idea is to start with  primitives in the DSL and build up increasingly more complex programs by combining existing terms via DSL constructs.  To scale up this approach to non-trivial synthesis tasks, PBE techniques based on bottom-up enumeration leverage the concept of \emph{observational equivalence}: Programs that produce the same output on the given set of input examples are effectively identical (at least for PBE purposes); hence, it suffices to keep only one representative of such observationally equivalent programs. For example, synthesis techniques based on \emph{finite tree automata (FTA)} leverage this observational equivalence idea to build a compact version space  representing the set of all programs consistent with the given input-output examples.

While this observational equivalence idea can significantly reduce the search space in some domains, it is not as effective in scenarios where few programs share the same (relevant) input-output behavior. As an example, consider the  \emph{inverse CSG\footnote{Constructive Solid Geometry} problem}, where the goal is to ``de-compile'' a complex geometric shape into a set of geometric operations that were used to construct it in a computer aided design (CAD) system
~\cite{Du2018,WillisPLCDLSM21}. While this problem can be framed as a PBE task~\cite{Du2018}, it is difficult to tackle this problem using domain-agnostic PBE engines because few programs  produce \emph{exactly} the same image. However, even in such domains, there are often many programs that have very similar, but not exactly identical, input-output behaviors. This observation motivates the following question: Can we relax the observational equivalence criterion and develop a synthesis algorithm that exploits the \emph{semantic similarity} between  different programs?

In this paper, we answer this question affirmatively and present a new  PBE algorithm called \emph{metric program synthesis} that can be applied in many  domains. Our method relaxes the standard \emph{observational equivalence} criterion to a weaker one called \emph{observational similarity} and groups together programs that produce \emph{similar} outputs on the same input. In particular, given a distance metric $\delta$, our method clusters programs into the same equivalence class if their output is within an $\epsilon$ radius with respect to $\delta$. Thus, in domains like inverse CSG where few DSL programs are observationally equivalent but many  have \emph{similar} input-output behaviors, such distance-based clustering can reduce the search space much more substantially than existing techniques.

To exploit observational similarity, our metric program synthesis approach proceeds in two phases: First, it performs bottom-up enumerative synthesis to build a \emph{version space}~\cite{vs} that compactly represents all programs up to some fixed AST depth. During  bottom-up enumeration, it clusters programs into equivalence classes using the provided distance metric, keeping one representative of each equivalence class. However,  due to the use of distance-based clustering, the generated version space is \emph{approximate}: it contains many programs that are incorrect but \emph{close to} being correct.  To deal with this difficulty, our method combines version space construction with a second \emph{local search} step: Starting with a program $P$ whose  output is close to the goal, it performs hill-climbing search to find a syntactic perturbation $P'$ of $P$ that has the intended input-output behavior. Because it is often possible to find \emph{syntactically} similar
representations of \emph{semantically} similar programs in many domains, this combination of approximate version space construction
with local program repair makes our approach effective.

We have implemented the proposed approach in a tool called \name{} and evaluate it on three different domains considered in prior work, namely (1) inverse CSG~\cite{Du2018}, (2) regular expression synthesis~\cite{alpha-regex,regel}, and (3) tower building~\cite{dreamcoder}. Our evaluation shows that \name{} is competitive with synthesizers designed/trained for these domains and that it  outperforms other domain-agnostic synthesizers based on the observational equivalence idea.

To summarize, this paper makes the following key contributions:
\begin{itemize}[leftmargin=*]
	\item We introduce  \emph{metric program synthesis}, a new synthesis approach  that exploits the semantic and syntactic similarity of programs in the search space.
	\item We show how to use distance metrics to perform effective clustering, ranking, and repair of programs explored during synthesis.
	\item We implement our technique in a new tool called \name{} and evaluate it in three different application domains, comparing \name to several relevant baselines.
\end{itemize}

\section{Overview}\label{sec:overview}

In this section, we work through an example from the inverse CSG domain that illustrates the key ideas in our algorithm.

\begin{figure}
	\begin{subfigure}[b]{0.3\textwidth}
		\centering
		\includegraphics[width=\textwidth]{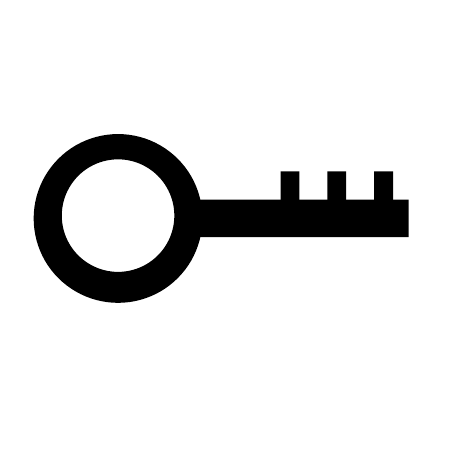}
		\caption{The input image.}\label{fig:example-pic}
	\end{subfigure}
	\begin{subfigure}[b]{0.3\textwidth}
		\centering
		\includegraphics[width=\textwidth]{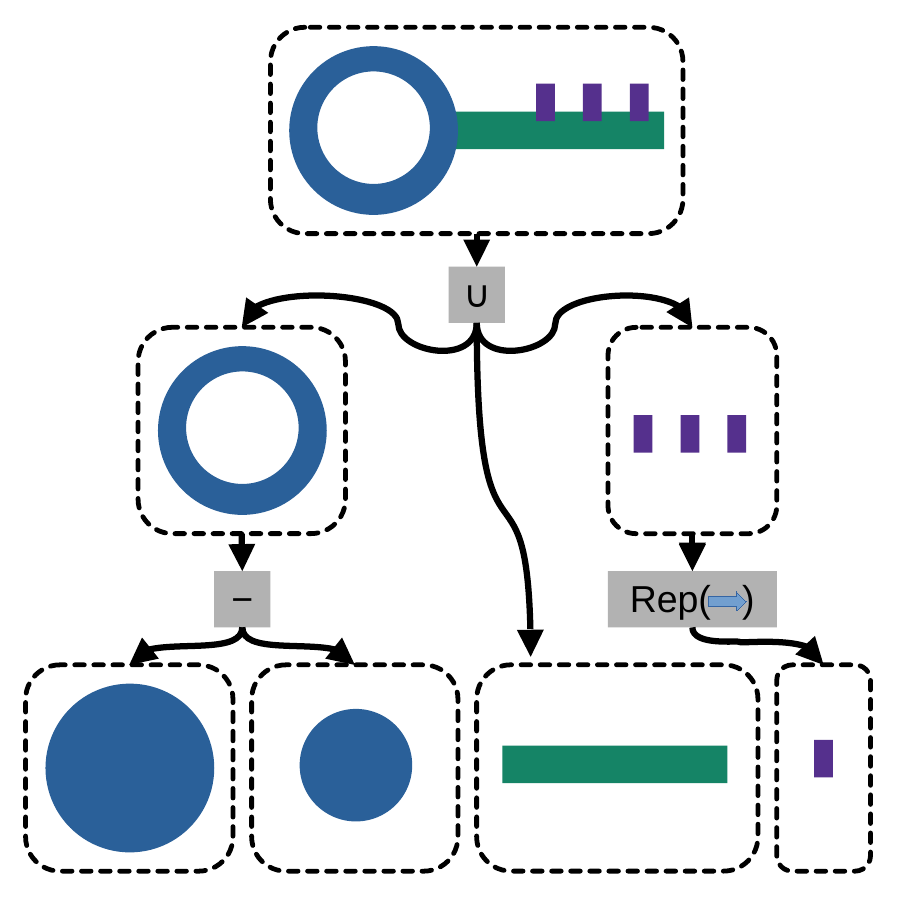}
		\caption{The solution, illustrated.}
	\end{subfigure}
	\begin{subfigure}[b]{0.3\textwidth}
		\begin{align*}
			 & (\texttt{Circle}(4, 8, 4) - \texttt{Circle}(4, 8, 3))     \\
			 & \cup\texttt{Rect}(7, 7, 15, 9)                            \\
			 & \cup \texttt{Repeat}(\texttt{Rect}(10, 9, 11, 10), 2, 0, 3)
		\end{align*}
		\caption{Program text of the solution.}\label{fig:example-text}
	\end{subfigure}
	\caption{An example CSG problem.}\label{fig:example}
\end{figure}

Consider the  picture of a key shown in \Cref{fig:example}. To a human,  it is  clear that this image contains three important component pieces: circles to make up the handle of the key, a rectangle for the shaft, and evenly spaced rectangles for the teeth. We can write a program that generates this key in a simple  DSL that includes primitive circles and rectangles as well as union, difference, and repetition operators. 
In particular, \Cref{fig:example-text} shows a program that can generate this picture. The program composes the three shapes: a hollow circle for the handle of the key, a rectangle for the shaft, and three small, evenly spaced rectangles for the teeth. The hollow circle is constructed by subtracting a small circle from a larger one: \[ \texttt{Circle}(x, y, r) - \texttt{Circle}(x', y', r') \]
The evenly spaced teeth are constructed by replicating a small rectangle three times: \[ \texttt{Repeat}(\texttt{Rect}(x, y, x', y'), dx, dy, 3) \] Circles are specified by a center point and radius, and rectangles are specified by their lower left and upper right corners.

Suppose that our goal  is to synthesize the program in \Cref{fig:example-text} given \emph{just} the picture in \Cref{fig:example-pic}. As mentioned in Section~\ref{sec:intro}, one standard approach is to perform bottom-up search over programs in the DSL: that is, create programs by composing together smaller terms, but discarding those that create the same shape as a previously-encountered one.
This approach---known as bottom-up enumeration with equivalence reduction~\cite{transit}---is a simple but powerful domain-agnostic synthesis algorithm that works well in many domains.

However, the problem  is that the inverse CSG domain is full of programs that are \emph{similar}, but not identical. To see why, consider the two circles that make up the handle of the key. If the outer circle was slightly larger or slightly shifted, it would still be clear to us that the  image is only slightly perturbed. We would be able to fix the program by locally improving it --- i.e., shifting the circle back into place by changing its parameters. It should not be necessary to retain both programs in the search space, since it is straightforward to transform one into the other. However, we cannot use equivalence reduction to group these two programs together, even though our intuition tells us that they should be nearly interchangeable.

To synthesize the figure above, our algorithm proceeds in two phases: It first performs coarse-grained search to look for a program $P$ that is \emph{close to} matching the target image. Then, in the second phase, it applies perturbations to $P$ in order to find a repair that \emph{exactly} matches the given image. We now explain these two phases in more detail.

\paragraph{Global coarse-grained search:} The first phase of our algorithm is based on bottom-up search and, like prior work~\cite{blaze}, it builds a data structure that compactly represents a large space of  programs. In particular, we represent the space of programs using a variant of a finite tree automaton (FTA) called an \emph{approximate finite tree automaton} (XFTA), which is described in detail in \cref{sec:xfta}.  The key idea behind an XFTA is to group together values that are semantically similar: in the CSG context, this means that images that are sufficiently similar to each other are represented using the same state in the automaton.


Our method constructs such an approximate tree automaton in three phases, namely \emph{expansion}, \emph{grouping}, and \emph{ranking}. In the expansion phase, operators are applied to sub-programs to create new candidate programs. For our running example, the first expansion step generates the set of primitive shapes. Later expansions compose shapes together using Boolean operators and the looping operator \texttt{Repeat} to create images of increasing complexity, but many of the images generated during the expansion phase are similar to each other.

In the grouping phase, images are put into clusters. Each cluster has a center $c$ and a radius $\epsilon$ such that every image in the cluster is within distance $\epsilon$ of $c$. Although every image in the cluster is retained as part of the search space, only the center of the cluster participates in further expansion steps. This clustering phase is essentially a relaxed version of equivalence reduction.

Finally, in the ranking phase, the $w$ clusters that are closest to the goal image are retained. This focuses the search on the programs that are likely to produce the goal. After ranking, the top $w$ clusters are inserted as new states into the XFTA, and the operators that produced each state in the cluster are inserted as edges.

When the forward search terminates, the XFTA represents a space of programs that are close to the target image, meaning that they produce visually similar images. However, at this point, a new difficulty presents itself: If we only clustered programs that had equivalent behavior, we could simply check whether the target image is present in the XFTA.\@ If it is, then the corresponding program can be extracted from the automaton and the synthesis task is complete. But, because we cluster programs that  have similar but not the same behavior, we may complete XFTA construction and find that the target image is not present in the automaton, even though that there are several images that are \emph{close} to the goal. To address this issue, our method performs a second level of \emph{local search}.

\paragraph{Local fine-grained search:} The local search proceeds in two phases: first, it extracts a candidate program from the XFTA; then, it attempts to repair the candidate.

Since each node in the XFTA represents a (possibly) exponentially large set of programs, we use a greedy algorithm to select a program from this set  rather than attempting to search over it. Starting at an accepting state of the automaton, the program extractor selects the incoming edge  that produces the closest image to the goal. Selecting an edge determines the root operator of the candidate program and the program sets from which to select arguments to that operator. Extraction proceeds recursively, always minimizing the distance between the overall candidate program and the goal.

\begin{figure}
	\centering
	\begin{subfigure}[b]{0.7\textwidth}
		\includegraphics[width=\textwidth]{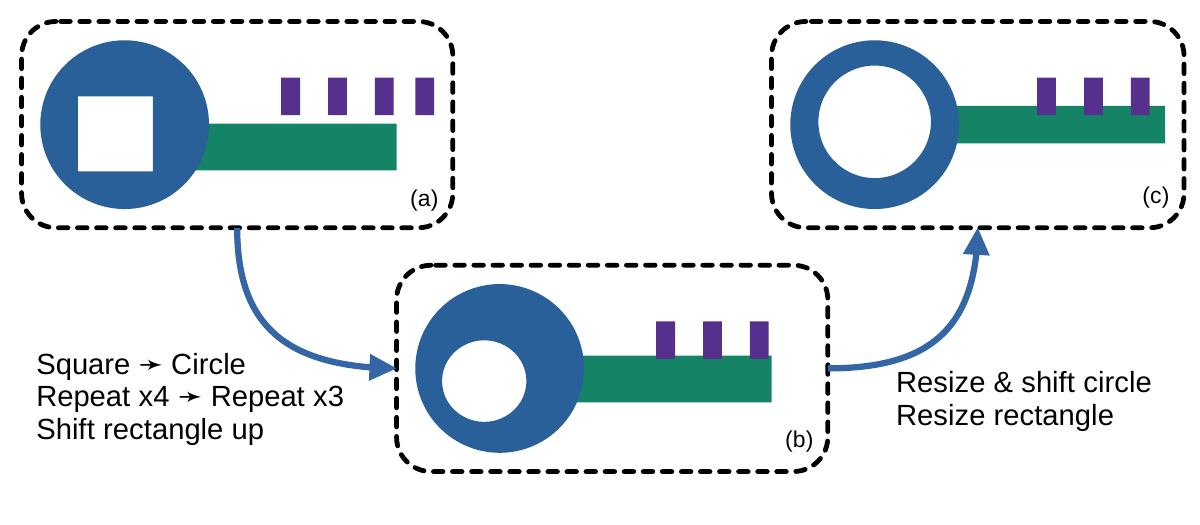}
		\phantomcaption\label{fig:local-search-begin}
	\end{subfigure}
	\begin{subfigure}{0\textwidth}
		\phantomcaption
	\end{subfigure}
	\begin{subfigure}{0\textwidth}
		\phantomcaption\label{fig:local-search-end}
	\end{subfigure}
	\vspace{-4ex}\caption{A sketch of the local search process for the key example.}\label{fig:local-search}
\end{figure}


When a candidate program has been extracted, we attempt to \emph{repair} it by applying syntactic rewrites. The sequence of rewrites is chosen using a form of tabu search~\cite{Glover1998} and is guided by the distance from the candidate program to the goal. At each step of the repair process, we consider the set of programs that can be obtained by applying a single rewrite rule to the current program and choose the one that is closest to the goal. This process continues until the desired program is found or until a maximum number of rewrites have been applied.

\Cref{fig:local-search} gives a high-level view of this repair process. Starting from a candidate program whose output is similar to the input image, the repair process applies rewrites such as changing squares to circles or incrementing/decrementing numeric parameters. Because  each rewrite gets us closer to the target image, the local search  can often quickly converge to a program that produces exactly the target image.  For instance, using our local search algorithm, we can obtain the program that produces the image from \Cref{fig:local-search-end} starting from the program for generating the image in \Cref{fig:local-search-begin}.


\section{Metric Program Synthesis Algorithm}\label{sec:algorithm}

In this section, we describe our proposed metric program synthesis algorithm. Given a domain-specific language $L$ and a set of of input-output examples of the form $\{ (I_1, O_1), \ldots, (I_n, O_n)\} $, the goal of our method is to synthesize a program $P$ in language $L$ such that $\forall i \in [1, n]. \ \llbracket P \rrbracket(I_i) = O_i$, meaning that evaluating $P$ on $I_i$ yields $O_i$ for every input-output example.

The rest of this section is organized as follows: First, because our method builds on bottom-up synthesis using finite tree automata, we start with some preliminary information on FTAs in Section~\ref{sec:fta}. Then, in Section~\ref{sec:alg_main}, we give an overview of our top-level synthesis algorithm, followed by discussions of its three  sub-procedures in Sections~\ref{sec:xfta}-\ref{sec:local}.

\subsection{Background on Synthesis using FTAs}\label{sec:fta}
Our synthesis algorithm builds on prior work on synthesis using finite tree automata  (FTA).
At a high level, an FTA is a generalization of a DFA from words to trees. In particular, just as a DFA accepts words, an FTA recognizes trees. More formally, FTAs are defined as follows:

\begin{definition} {\bf (FTA)}
	A bottom-up finite tree automaton (FTA) over alphabet $\alphabet$ is a tuple $\fta = (\ftastates, \finalstates, \transitions)$ where $\ftastates$ is the set of states, $\finalstates \subseteq \ftastates$ are the final states, and $\transitions$ is a set of transitions of the form
	$\ell(\ftastate_1, \dots, \ftastate_n) \rightarrow \ftastate$
	where $\ftastate, \ftastate_1, \dots, \ftastate_n \in \ftastates$ and $\ell \in \alphabet$.
\end{definition}

Intuitively, FTAs are useful in synthesis because they can compactly encode a set of programs, represented in terms of their abstract syntax tree~\cite{blaze,dace,burst}. In particular, when used in the context of synthesis, states of the FTA correspond to values (e.g., integers), and the alphabet corresponds to the set of DSL operators (e.g., $+, \times$). Final states are marked based on the specification, and transitions model the semantics of the underlying DSL.\@ For instance, in a language with a negation operator $\neg$, transitions  $\neg(0) \rightarrow 1$ and $\neg(1) \rightarrow 0$ express the semantics of  negation.

We can view  terms over an alphabet $\alphabet$ as  trees of the form  $T = (n, V, E)$ where $n$ is the root node, $V$ is a set of labeled vertices, and $E$ is the set of edges.
A term $T$ is said to be accepted by an FTA if  $T$ to can be rewritten to some state $q \in Q_f$ using transitions $\transitions$.
Finally, the language of a tree automaton $\fta$ is  denoted as $\mathcal{L}(\fta)$ and consists of the set of all terms accepted by~$\fta$.

Given a specification $\varphi$, the idea behind FTA-based synthesis is to construct an FTA whose language is the set of all programs satisfying $\varphi$. In particular, FTAs can be used to solve the programming-by-example as follows: For each input-output example $(I, O)$, the idea is to start with a state representing $I$ and construct new states and transitions by applying the DSL operators. For example, given FTA states representing integers $1$ and $2$ and a $+$ operator in the DSL, we generate a new FTA state representing $3$  using the transition $+(1,2) \rightarrow 3$. This process of adding new states and transitions to the FTA continues until either there are no more states to be added or a pre-defined bound on the number of states or transitions has been reached. The final state of the FTA corresponds to the output example $O$, and the standard  intersection operator is used to generate an FTA whose language includes programs that are consistent with \emph{all} input-output examples.

As this discussion makes clear, such an FTA-based approach can compactly represent the version space if there are many DSL programs that share the \emph{same} input-output behavior (because such programs lead to the same FTA state). However,  FTA-based synthesis may not  scale in application domains that do not have this property. Prior work on FTA-based synthesis has tried to tackle this problem using abstract interpretation and abstraction refinement~\cite{blaze}: in that setting, FTA states correspond to \emph{abstract} rather than \emph{concrete} values, and transitions are constructed using the \emph{abstract}, rather than the concrete, semantics of the DSL.  Of course, since such an \emph{abstract FTA} over-approximates the set of programs consistent with the specification, one needs to perform abstraction refinement to iteratively rule out spurious programs from the language of the FTA.  While this so-called {\sc Syngar} approach has proven to be effective in some domains like tensor manipulations~\cite{blaze},  such abstractions are not universally easy to construct. For example, we have found the inverse CSG domain to \emph{not} be particularly friendly for such an abstract interpretation approach. Our metric-based synthesis algorithm is an attempt to solve this problem in a different way using distance metrics rather than abstract domains.

\subsection{Overview of Synthesis Algorithm}\label{sec:alg_main}
As mentioned earlier, the key idea behind metric-based synthesis is to relax the \emph{observational equivalence} criterion into \emph{observational similarity} by using a distance metric. More formally, we define observational similarity as follows:

\begin{definition}{\bf (Similarity)}
	Two values  $v$ and $v'$ are \emph{similar}, denoted $v \simeq_{\epsilon} v'$ if they are within $\epsilon$ of each other, according to a distance metric $\delta$: \[v \simeq_{\epsilon} v' \Leftrightarrow \delta(v, v') \leq \epsilon.\]
\end{definition}

\begin{algorithm}[t]
	\begin{algorithmic}[1]\onehalfspacing
		\Require $L$ is a language, $I$  and $O$  are the input and output examples respectively, $c_{max}$ is the maximum program size to consider when constructing the XFTA, $w$ is the beam width, $\delta$ is a distance metric between values, $\epsilon$ is the threshold for clustering.
		\Ensure On success, returns a program $p$ where $\eval{p}=O$. On failure, returns $\bot$.
		\Procedure{MetricSynth}{$L, I, O, c_{max}, w, \delta, \epsilon$}
		\State $\fta <- \Call{ConstructXFTA}{L, I, O, c_{max}, w, \delta, \epsilon}$\label{l:coarse}
		\For{$P \in \Call{Extract}{\fta, I, O, q, \delta}$}
		\State $P <- \Call{Repair}{I, O, \delta, P}$
		\If{${P} \neq \bot$}
		\State \Return $P$
		\EndIf
		\EndFor\label{l:fine-end}
		\State \Return $\bot$
		\EndProcedure
	\end{algorithmic}

	\caption{Metric synthesis algorithm.}\label{alg:metric}
\end{algorithm}


The main idea behind our synthesis algorithm is to construct an approximate version space by clustering together  values that are similar. Just as the {\sc Syngar} idea of ~\citep{blaze} groups together values based on an \emph{abstract} notion of  observational equivalence, our method groups together values based on this notion of \emph{similarity} and constructs a so-called \emph{approximate FTA (XFTA)} representing programs that produce values close to the desired output.

Our top-level metric program synthesis approach is presented in Algorithm~\ref{alg:metric} and is parameterized over a (1) {distance metric} $\delta$, (2) radius $\epsilon$, and (3) domain-specific language $L$. At a high level, this algorithm consists of three steps:

\begin{enumerate}[leftmargin=*]
	\item {\bf XFTA construction:}  First, {\sc MetricSynth} constructs an FTA that represents a space of programs that produce values close to the goal (line 2 of Algorithm~\ref{alg:metric}). However, because this FTA is constructed by grouping similar values together, a program excepted by this automaton does not necessarily satisfy the specification.
	\item {\bf Program extraction:} To deal with the approximation introduced by clustering, the algorithm enters a loop in which it repeatedly extracts programs from the FTA using the call to {\sc Extract} at line 3. The goal of {\sc Extract} is to find a program in the language of the FTA that produces a value that is sufficiently close to the target.
	\item {\bf Program repair:} Because the extracted program does not satisfy the input-output examples in the general case, the {\sc Repair} procedure (invoked at line 4) tries to find a  syntactic perturbation of $P$ that exactly satisfies the input-output examples. As we discuss in more detail in Section~\ref{sec:local}, the repair procedure is based on rewrite rules and performs a form of tabu search, using the distance metric as a guiding heuristic.
\end{enumerate}

In the following subsections, we discuss the {\sc ConstructXFTA}, {\sc Extract}, and {\sc Repair} procedures in more detail.

\subsection{Approximate FTA Construction}\label{sec:xfta}
\begin{algorithm}[!t]
	\begin{algorithmic}[1]\onehalfspacing
		\Require $\Sigma$ is a set of operators, all other parameters are the same as in \Cref{alg:metric}. $k$ is a hyper-parameter that determines the number of states that the automaton should accept.
		\Ensure Returns an XFTA.
		\Procedure{ConstructXFTA}{$\Sigma, I, O, c_{max}, w, \delta, \epsilon$}
		\State $Q <- I, \Delta <- \emptyset$
		\For{$1 \leq c \leq c_{max}$}
		\State $\Delta_\emph{frontier} <- \left\{ \ell(q_1, \dots, q_n) -> q ~\Big|~~
			\begin{aligned}
				 & \ell \in \Sigma, \quad \{q_1, \dots, q_n\} \subseteq Q,\quad \eval{\ell(q_1, \dots, q_n)} = q \\
			\end{aligned} \right\}$
		\State $(Q_c, \Delta_\emph{c}) <- \Call{Cluster}{\Delta_\emph{frontier}, \delta, \epsilon}$
		\State $Q' \gets \Call{TopK}{Q_c, \delta(O), w } $
		\State $Q \gets Q \cup Q'$
		\State $\Delta \gets \Delta \cup \{ (\ell(q_1, \dots, q_n) -> q) \  | \ q \in Q',  (\ell(q_1, \dots, q_n) -> q) \in \Delta_c \} $
		\EndFor
		\State $Q_{f} <- \Call{TopK}{Q, \delta(O), k}$
		\State \Return $(Q, Q_{f}, \Delta)$
		\EndProcedure
	\end{algorithmic}
	\caption{Algorithm for constructing an approximate FTA.}\label{alg:coarse}
\end{algorithm}

\Cref{alg:coarse} shows our technique for constructing an approximate FTA for a given set of input-output examples. At a high level, this algorithm builds programs in a bottom-up fashion, clustering together those programs that produce similar values on the same input. In order to ensure that the algorithm terminates, it only builds programs up to some fixed depth controlled by the hyper-parameter $c_\emph{max}$.

In more detail, {\sc ConstructXFTA}  adds new automaton states and transitions (initialized to $I$ and $\emptyset$ respectively) in each iteration of the while loop. In particular, for each (n-ary) DSL operator $\ell$ and existing states $q_1, \ldots, q_n$, it obtains a new \emph{frontier} of  candidate transitions $\Delta_\emph{frontier}$ by evaluating $\ell(q_1, \ldots, q_n)$. The construction of this frontier corresponds to the \emph{expansion phase} mentioned in Section~\ref{sec:overview}.

\begin{algorithm}[t]
	\begin{algorithmic}[1]\onehalfspacing
		\Require $\Delta$ is a set of FTA transitions, all other parameters are the same as in \Cref{alg:metric}.
		\Ensure Returns a set of FTA transitions.
		\Procedure{Cluster}{$\Delta, \delta, \epsilon$}
		\State $Q' \gets \emptyset$ \Comment{New (clustered) states}
		\State $\Delta' \gets \emptyset$ \Comment{New (clustered) transitions}
		\For{$(\ell(q_1, \dots, q_n) -> q) \in \Delta$}
		\State $close <- \{q_{center} \in Q' ~|~ q_{center} \simeq_{\epsilon} q\}$
		\If{$close = \emptyset$}
		\State $Q' \gets Q' \cup \{ q \}$
		\State $\Delta' \gets \Delta' \cup \{\ell(q_1, \dots, q_n) -> q\} $
		\Else
		\State $\Delta_\emph{new} \gets \{ \ell(q_1, \dots, q_n) -> q'  \ | \ q' \in close \} $
		\State $\Delta' \gets \Delta' \cup \Delta_\emph{new}$
		\EndIf
		\EndFor
		\State \Return $(Q', \Delta')$
		\EndProcedure
	\end{algorithmic}

	\caption{Greedy algorithm for clustering states.}\label{alg:clustering}
\end{algorithm}

In general, the expansion phase can result in a very large number of new states, making XFTA construction prohibitively expensive. Thus, in the next \emph{clustering phase} (line 5 of \Cref{alg:coarse}), the algorithm groups similar states introduced by expansion into a single state as shown in \Cref{alg:clustering}. In this context, standard clustering algorithms like k-means are not suitable because they fix the number of clusters but allow the radius of each cluster to be arbitrarily large. In contrast, we would like to minimize the number of clusters while ensuring that the radius of each cluster is bounded. Hence, we use the {\sc Cluster} procedure from \Cref{alg:clustering} to generate a set of clusters where each state is within some $\epsilon$ distance from the center of a cluster. To do so, \Cref{alg:clustering} iterates over the new states $q$ in the frontier and starts a new cluster for $q$ if none of the previous frontier states are within $\epsilon$ of $q$ (lines 6--8 in \Cref{alg:clustering}). Otherwise, $q$ is added to an existing cluster (lines 9--11 in \Cref{alg:clustering}).  Furthermore, for each new transition $\ell(q_1, \ldots, q_n) \rightarrow q$ of the frontier,  clustering produces new transitions of the form $\ell(q_1, \ldots, q_n) \rightarrow q_c$ where $q_c$ is the center of a cluster that $q$ belongs to. Hence, clustering produces a new set of states $Q_c$ and a new set of transitions $\Delta_c$ to add to the automaton, as shown in line 5 of \Cref{alg:coarse}.

The final component of XFTA construction is the \emph{ranking phase}, which corresponds to lines 6--8 of \Cref{alg:coarse}. Even after  clustering, the automaton might end up with a prohibitively large number of new states, so {\sc ConstructXFTA} only keeps the top $w$ clusters in terms of their distance to the goal. Thus, in each iteration, \Cref{alg:coarse} only ends up adding $w$ new states to the automaton, similar to beam search.

\begin{algorithm}[!t]
	\begin{algorithmic}[1]\onehalfspacing
		\Require $\fta$ is an XFTA;\@ all other parameters are the same as in \Cref{alg:metric}.
		\Ensure Yields program terms that are accepted by $\fta$.
		\Procedure{Extract}{$\fta, I, O, \delta$}
		\State $Q_f <- \Call{FinalStates}{\fta}$
		\State Sort $Q_{f}$ by $\delta(O)$ increasing.\label{l:fine-start}
		\For{$q_f \in Q_{f}$}
		\State $\Delta <- \textsc{Transitions}(\fta)$
		\State $\Delta_{root} <- \{ (\ell(q_1, \dots, q_{n}) -> q_f) ~|~ (\ell(q_1, \dots, q_{n}) -> q_f) \in \Delta \}$
		\State \textbf{yield} \Call{ExtractTerm}{$\Delta_{root}, I, \delta(O)$}
		\EndFor
		\EndProcedure
		\item[]
		\Require $\Delta$ is a set of FTA transitions, $\delta$ is a distance metric.
		\Ensure Returns a program term.
		\Procedure{ExtractTerm}{$\Delta, I, \delta$}
		\State Let $(\ell(q_1, \dots, q_n) -> q) \in \Delta$ be a transition where $q$ minimizes $\delta(q)$
		\For{$1 \leq i \leq n$}\Comment{Extract a program for each argument to $\ell$.}
		\State $\delta_{i} <- \lambda q.\ \delta(\eval{\ell(q_1', \dots, q_{i-1}', q, q_{i+1}, \dots, q_{n})})$
		\State $p_{i} <- \Call{ExtractTerm}{\Delta, I, \delta_{i}}$
		\State $q_{i}' <- \eval{p_{i}}(I)$
		\EndFor
		\State \Return $\ell(p_{1}, \dots, p_{n})$
		\EndProcedure
	\end{algorithmic}

	\caption{Algorithm for extracting programs from an XFTA.}\label{alg:reverse}
\end{algorithm}


\subsection{Extracting Programs from XFTA}\label{sec:extract}

We now turn our attention to the {\sc Extract} procedure for picking a program that is accepted by our approximate FTA.\@ Recall that programs accepted by the XFTA are not necessarily consistent with the input-output examples due to  clustering. Furthermore, two programs $P, P'$ that are accepted by the XFTA need not be equally close to the goal state; for example, $\eval{P}(I)$ might be much closer to $O$ than $\eval{P'}(I)$ with respect to the distance metric $\delta$. Ideally, we would like to find the best program that is accepted by the FTA (in terms of its proximity to the goal); however, this can be prohibitively expensive, as the automaton (potentially) represents an exponential space of programs. Thus, rather than finding the best program accepted by the automaton, our {\sc Extract} procedure uses a greedy approach to yield a sequence of ``good enough'' programs in a computationally tractable way.

The high-level idea behind {\sc Extract} is to recursively construct a program starting from the specified final state $q_f$ via the call to the recursive procedure {\sc ExtractTerm}. At every step, the algorithm picks a transition $\ell(q_1, \ldots, q_n) \rightarrow q$ whose output  minimizes the distance from the goal and then recursively constructs the arguments $p_1, \ldots, p_n$ of $\ell$. Note that this algorithm is greedy in the sense that it tries to find a single  operator that minimizes the distance from the goal rather than a sequence of operators (i.e., the whole program). Hence, among the programs accepted by $\fta$, there is no guarantee that {\sc Extract} will return the globally optimum one.

\subsection{Distance-Guided Program Repair}\label{sec:local}

The final part of our synthesis algorithm (\textsc{Repair}) takes the program that was extracted from the XFTA and attempts to repair it by applying syntactic rewrite rules. In particular, given a program $P$ that is close to the goal, {\sc Repair} tries to find a program $P'$ that is (1) syntactically close to $P$ and (2) correct with respect to the input-output examples (i.e., $\eval{P}(I) = O)$.

Our \textsc{Repair} procedure  is parameterized by a set of rewrite rules $R$ of the form $t -> s$. We say that a program $P$ can be rewritten into $P'$ if there is a rule $r = (t -> s) \in R$ and a substitution $\sigma$ such that $P = \sigma t$ and $P' = \sigma s$. We denote the application of rewrite rule $r$ to $P$  as $P ->_{r} P'$.

The \textsc{Repair} procedure is presented in \Cref{alg:repair} and applies goal-directed rewriting to the candidate program, using the distance function $\delta$ to guide the search. In particular, it starts with the input program $P$ and iteratively applies a rewrite rule until  either a correct program is found or a bound $n$ on the number of rewrite rules is reached. In each iteration of the loop (lines 3--8), it first generates a set of new candidate programs (called \emph{neighbors}) by applying a rewrite rule to $P$ and (greedily) picks the program $P'$ that minimizes the distance $\delta(O, \eval{P'}(I))$. In the next iteration, the new program $P'$ is used as the seed for applying rewrite rules.

\begin{algorithm}[!t]
	\begin{algorithmic}[1]\onehalfspacing
		\Require $I, O$ are the input output examples, $\delta$ is a distance metric, $P$ is a program. There are also two hyperparameters: $n$ is the maximum number of rewrites to perform, and $R$ is a set of rewriting rules.
		\Ensure Returns a program $P'$ such that $\eval{P'}(I) = O$ or returns $\bot$.
		\Procedure{Repair}{$I, O, \delta, P$}
		\State $S \gets \emptyset$
		\While{$i < n$}
		\State $neighbors <- \{ P' ~|~ P ->_{r} P', r \in R \} - S$ 
		\State $P <- \argmin_{p \in neighbors} \delta(O, \eval{p}(i))$
		\If{$\eval{P}(I) = O$}
		\State \Return $P$
		\EndIf
		\State $S \gets S \cup P$
		\EndWhile
		\State \Return $\bot$
		\EndProcedure
	\end{algorithmic}
	\caption{Algorithm for repairing a program.}\label{alg:repair}
\end{algorithm}


Note that our {\sc Repair} procedure utilizes a (bounded) set $S$ to avoid getting stuck in local minima, as is done in tabu search~\cite{Glover1998}. In particular, $S$ contains the most recently explored $k$ programs, and, when applying a rewrite rule, the {\sc Repair} procedure avoids generating any program  in set $S$.

\section{Instantiating Metric Synthesis in Application Domains}\label{sec:inst}
The {\sc MetricSynth} algorithm presented so far is DSL-agnostic, so it can be instantiated in different application domains. In this section, we discuss how we instantiate the metric synthesis algorithm in the (1) inverse CSG, (2) regular expression synthesis, and (3) tower-building domains.

\subsection{Instantiation for inverse CSG}\label{sec:inst_cad}

The \emph{inverse CSG problem}  aims  to ``de-compile'' a complex geometric shape into a set of geometric operations that were used to construct it. We now describe how to instantiate our metric program synthesis framework to solve the inverse CSG problem.

\subsubsection{Domain-Specific Language}

\begin{figure}
	\begin{align*}
		E\ :=\ \texttt{Circle}(x, y, r) ~|~ \texttt{Rect}(x_1, y_1, x_2, y_2) ~|~ E \cup E ~|~ E - E ~|~ \texttt{Repeat}(E, x, y, c)
	\end{align*}
	\caption{The syntax of CSG programs.}\label{fig:syntax}
\end{figure}


\Cref{fig:syntax} shows the syntax of the domain-specific language that we use for the inverse CSG problem. This DSL includes two primitive shapes, namely circles and rectangles. Circles are represented by a center coordinate and a radius. Rectangles are axis-aligned and are represented by the coordinates of the lower left and upper right corners. These primitive shapes can be combined using union, difference, and repeat operators. In particular,
$\texttt{Repeat}(E, x, y, c)$ takes an image $E$, a translation vector $v = (x, y)$, and a count $c$, and  it produces the union of $E$ repeated $c$ times, translated by $v$. For example, we have: $$\texttt{Repeat}(\texttt{Circle}(v, r), v', 2) = \texttt{Circle}(v, r) \cup \texttt{Circle}(v + v', r)$$
\texttt{Repeat} allows programs with repeating patterns to be expressed compactly, which also makes these programs easier to synthesize.

\Cref{fig:semantics} presents the semantics of our  CSG DSL using an {\sc Eval} procedure, which takes as input a program and produces a bitmap image. Specifically, we represent a bitmap image as a mapping from each pixel $(u, v)$ to a Boolean indicating whether that pixel is filled or not, or equivalently as the set of all pixels that evaluate to true.   Given a program $P$ in our DSL, we use the notation $\eval{P}$ to denote the bitmap image produced by $P$.

\begin{figure}
	\begin{align*}
		M                                                & = \{(u, v) ~|~ 0 \leq u < x_{max}, 0 \leq v < y_{max}\}                                                  \\
		\textsc{Eval}(\texttt{Circle}(x, y, r))          & = \left\{(u, v) ~|~ \sqrt{{(x-u)}^2 + {(y-v)}^2} < r ~|~ (u, v) \in M \right\}                              \\
		\textsc{Eval}(\texttt{Rect}(x_1, y_1, x_2, y_2)) & = \left\{(u, v) ~|~ x_1 \leq u \land y_1 \leq v \land u \leq x_2 \land v \leq y_2 ~|~ (u, v) \in M \right\} \\
		\textsc{Eval}(e \cup e')                         & = \left\{(u, v) ~|~ \textsc{Eval}(e)[u, v] \lor \textsc{Eval}(e')[u, v] ~|~ (u, v) \in M \right\}           \\
		\textsc{Eval}(e - e')                            & = \left\{(u, v) ~|~ \textsc{Eval}(e)[u, v] \land \lnot \textsc{Eval}(e')[u, v] ~|~ (u, v) \in M \right\}    \\
		\textsc{Eval}(\texttt{Repeat}(e, x, y, c))       & = \left\{(u, v) ~|~ \bigvee_{0 \leq i < c} \textsc{Eval}(e)[u+ix,v+iy] ~|~ (u, v) \in M \right\}
	\end{align*}
	\caption{The semantics of CSG programs, given as an evaluation function.}\label{fig:semantics}
\end{figure}


\subsubsection{Synthesis Problem}

Given a bitmap image $B$, represented as a mapping from pixels $(x, y)$ to  Booleans, inverse CSG aims to synthesize a program $P$ such that:
\[
	\forall x, y \in \emph{Domain}(B). \ B(x, y) = \eval{P} (x, y)
\]

Viewed in this light, note that inverse CSG is exactly a programming-by-example (PBE) problem: We can think of the bitmap image as a set of I/O examples where each input example is a pixel $(x, y)$ and the output example is a Boolean.
However, a key difference from standard PBE is that the number of examples we need to deal with is quite large: for instance, in our evaluation, we use $32 \times 32$ bitmap images, so the number of I/O examples is 1024.

\subsubsection{Distance function}
\begin{figure}
	\centering
	\includegraphics{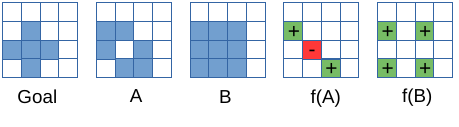}
	\caption{Illustration of the distance transformation $f_O$. Note that the Jaccard distance is $\delta_J(A, B) = \frac{1}{3}$, whereas $\delta_{Goal}(A, B) = \frac{3}{5}.$ This shows how the distance between images that are very close to the goal is magnified.}
	\label{fig:dist-illustr}
\end{figure}

A key component of our metric synthesis algorithm is the specific distance metric used. For the inverse CSG domain, we use a distance metric $\delta$ that is a slight modification of the standard Jaccard distance that takes into account the goal value. Specifically, we define the distance metric as follows:
\[\delta_O(q, q') = 1 - \frac{|f_O(q) \cap f_O(q')|}{|f_O(q) \cup f_O( q')|}\quad\text{where }f_O(q) = \{(x, y, b) ~|~ (x, y):b \in q, b \neq O[x, y]\}.\]
Intuitively, this distance considers only the pixels of the images $q$ and $q'$ that \emph{differ} from the goal image $O$. This is desirable because images that are very close to the goal are treated as further from each other than images that are far from the goal. Overall, this metric has the effect of making our algorithm more sensitive to differences in images that are close to the goal. We illustrate this effect in \Cref{fig:dist-illustr}.

\begin{theorem}
	$\delta_O$ is a metric on the set of images.
\end{theorem}
\begin{proof}
	Let $O$ be some goal value. Let $\delta_J(q, q') = 1 - \frac{|q\cap q'|}{|q \cup q'|}$ be the Jaccard distance, which is a metric on finite sets. We have $\delta_O(q, q') = \delta_J(f_O(q), f_O(q'))$ where $f_O$ (defined above) is a function $Image -> \mathbb{Z} \times \mathbb{Z} \times \mathbb{B}$.  Because an injective function $f$ from any set $S$ to a metric space $(M, \delta)$ gives a metric $\delta(f(x), f(x'))$ on $S$, we can show that $\delta_O$ is a metric by showing that $f$ is injective. To see why this is the case, note that we can define the inverse of $f_O$ as follows:
	\[f_O^{-1}(s) = \left\{(x, y):b' ~|~ (x, y):b \in O, b' = \lnot b\text{ if }(x, y, \lnot b) \in s\text{ else }b\right\}.\]
	Intuitively, where $f_O$ returns the differences between $q$ and $O$, $f_O^{-1}$ applies those differences to $O$ to obtain $q$.
\end{proof}

\subsubsection{Rewrite rules for repair.}
Recall that our \textsc{Repair} procedure is parameterized over a set of rewriting rules $R$. At a high level, the rewrite rules we use for the inverse CSG problem modify integers (within bounds) and transform squares into circles (and vice-versa). In more detail, our implementation utilizes the following rewrite rules:
\begin{align*}
	x                                         & -> x + 1                                     & \text{if }x\text{ is an integer and }x < x_{max}                           \\
	x                                         & -> x - 1                                     & \text{if }x\text{ is an integer and }x > 0                                 \\
	\texttt{Circle}(x, y, r)                  & -> \texttt{Rect}(x - r, y - r, x + r, y + r) &                                                                            \\
	\texttt{Rect}(x_{1}, y_{1}, x_{2}, y_{2}) & -> \texttt{Circle}(x_{1} + r, y_{1} + r, r)  & \text{where }r = \frac{x_{2}-x_{1}}{2}\text{ if }x_{2}-x_{1} = y_{2}-y_{1}
\end{align*}

\subsection{Instantiation for Regular Expressions}

Our second application domain of metric program synthesis is generating regular expressions from a set of positive and negative examples.

\subsubsection{Domain-Specific Language}

Our regular expression language (\Cref{fig:regex-syntax-semantics}) is similar to the one used in prior work~\cite{regel}. This DSL includes character classes, concatenation, repetition, optional matches, conjunction, disjunction, and negation. Character classes match a single character from a set; we use a set of single-character classes that includes all of the printable characters as well as multi-character classes for numbers, capital and lowercase letters, symbols, and vowels. $\texttt{Concat}(E, E')$ matches $E$ followed by $E'$. For example, $\texttt{Concat}(\texttt{<num>}, \texttt{<a>})$ matches the string ``0a.'' $\texttt{Repeat}(E, x)$ matches $x$ repetitions of $E$. As in the CSG DSL, we include the repetition operators to make repetitive programs more tractable to synthesize.

Programs in the regular expression DSL evaluate to a set of match locations in a string $s$, as shown in \Cref{fig:regex-syntax-semantics}. Since the semantics of regular expressions are fairly standard, we do not explain the semantics in detail.

\begin{figure}
	\centering
	\begin{align*}
		E := & \ \emptyset ~|~ C ~|~ \texttt{Concat}(E, E) ~|~ \texttt{Repeat}(E, x) ~|~ \texttt{RepeatRange}(E, x_1, x_2) ~|~ \texttt{RepeatAtLeast}(E, x) \\
		     & |~ \texttt{Optional}(E) ~|~ E \land E ~|~ E \lor E ~|~ \lnot E
	\end{align*}
	\begin{align*}
		\textsc{Eval}(C, s)                                 & = \{ (i, i+1) ~|~ 0 \leq i \leq |s|, s[i] \in C \}                                          \\
		\textsc{Eval}(\emptyset, s)                         & = \{ (i, i) ~|~ 0 \leq i \leq |s| \}                                                        \\
		\textsc{Eval}(\texttt{Concat}(E, E'), s)            & = \{ (i, k) ~|~ (i, j) \in \textsc{Eval}(E, s), (j', k) \in \textsc{Eval}(E', s), j = j' \} \\
		\textsc{Eval}(\texttt{Repeat}(E, 1), s)             & = \textsc{Eval}(E, s)                                                                     \\
		\textsc{Eval}(\texttt{Repeat}(E, x), s)             & = \textsc{Eval}(\texttt{Concat}(E, \texttt{Repeat}(E, x-1)), s)                           \\
		\textsc{Eval}(\texttt{RepeatRange}(E, x_1, x_2), s) & = \bigcup_{x_1 \leq i \leq x_2} \textsc{Eval}(\texttt{Repeat}(E, i), s)                   \\
		\textsc{Eval}(\texttt{RepeatAtLeast}(E, x), s)      & = \textsc{Eval}(\texttt{RepeatRange}(E, x, |s|), s)                                       \\
		\textsc{Eval}(\texttt{Optional}(E), s)              & = \textsc{Eval}(E \lor \emptyset, s)                                                      \\
		\textsc{Eval}(E \land E', s)                        & = \textsc{Eval}(E, s) \cap \textsc{Eval}(E', s)                                           \\
		\textsc{Eval}(E \lor E', s)                         & = \textsc{Eval}(E, s) \cup \textsc{Eval}(E', s)                                           \\
		\textsc{Eval}(\lnot E, s)                           & = \{ (i, j) ~|~ 0 \leq i \leq j \leq |s|, (i, j) \not \in \textsc{Eval}(E, s) \}
	\end{align*}
	\caption{Syntax and semantics of the regular expression DSL.}
	\label{fig:regex-syntax-semantics}
\end{figure}

\subsubsection{Synthesis Problem}
As in prior work~\cite{alpha-regex,regel}, we consider the problem of synthesizing regular expressions from a given set of positive and negative examples. Let $S^{+}$ be a set of positive examples  (strings), and let $S^{-}$ be a set of negative examples. Then, the synthesis problem is to generate a regular expression $E$ such that:
\[
	\forall s \in S^{+}. \ (0, |s|) \in {\textsc{Eval}}(E, s) \ \text{and}\ \forall s \in S^{-}. \ (0, |s|) \not \in {\textsc{Eval}}(E, s)
\]

However, prior work has shown that synthesizing the \emph{intended} regular expression just from positive and negative examples can be challenging if one only has access to few examples~\cite{regel}. For this reason, ~\cite{regel} has advocated using \emph{sketches} obtained from natural language descriptions. Following that work, we  consider a modified version of this problem where the regular expression needs to be a completion of the provided sketch in addition to satisfying the positive and negative examples. Specifically, our problem formulation additionally requires a sketch which is given in the form of a program with holes, where the holes may additionally contain constraints on the terms that must be used to fill the hole, as in~\cite{regel}.

\subsubsection{Distance Function}
We use a simple distance function that simply counts the number of examples that the regular expression matches.  In particular, we consider a program to match a positive example $s$ if $(0, |s|)$ is in the match set $m$. A program matches a negative example if $(0, |s|)$ is not in $m$.

\subsubsection{Rewrite rules}
Some of the constructs used in the regex DSL take integers as arguments. As in the inverse CSG instantiation, we include rewrite rules that transform integers. We also include a rule that transforms $\texttt{Rewrite}$ to $\texttt{RewriteRange}$, which allows programs that use \texttt{Rewrite} to be generalized to a range of repeat counts.

\begin{align*}
	x                      & -> x + 1                          & \text{if }x\text{ is an integer and }x < x_{max} \\
	x                      & -> x - 1                          & \text{if }x\text{ is an integer and }x > 0       \\
	\texttt{Repeat}(E, x) & -> \texttt{RepeatRange}(E, x, x)
\end{align*}

\subsection{Instantiation for Tower Building}

Our third application domain is the tower building task from prior work~\cite{dreamcoder,blended} that is  inspired by AI planning tasks. Given a set of blocks and target ``tower" (i.e. configuration of these blocks), this task aims to generate the desired tower by (programmatically) controlling a robot arm.

\subsubsection{Domain-specific Language}
\Cref{fig:tower-syntax-semantics} shows the syntax of the tower building DSL. Programs in this language control a robot arm which can move left and right along a horizontal track and can drop horizontal or vertical blocks. The state of the program includes the $x$-position of the arm and the list of dropped blocks.
In more detail, the DSL includes operators for dropping blocks, moving the robot arm, sequencing, and looping. It also includes the $\texttt{Embed}(E)$ operator, which executes $E$ and then restores the position of the arm. This operator gives the DSL a limited way to return to a previous state without needing to specify the movement needed to reset the arm.

The semantics of the DSL are shown in \Cref{fig:tower-syntax-semantics}. \texttt{DropH} and \texttt{DropV} both add a new (horizontal or vertical) block to the tower. The block will be placed on the highest block that is below the arm. The move operators both update the position of the arm; \texttt{MoveBefore} moves the arm and then executes $E$, while \texttt{MoveAfter} moves the arm after evaluating $E$. \texttt{Embed} gives the language a degree of modularity. It executes $E$ and then resets the arm to wherever it was before. Note that these semantics correspond to an idealized physics model where blocks fall in a perfectly straight line until they land on top of another block.

\begin{figure}
	\centering
	\begin{align*}
		E := \texttt{DropH} ~|~ \texttt{DropV} ~|~ \texttt{MoveBefore}(E, x) ~|~ \texttt{MoveAfter}(E, x) ~|~ E; E ~|~ \texttt{Loop}(x, E) ~|~ \texttt{Embed}(E)
	\end{align*}
	\begin{align*}
		\textsc{Eval}(\texttt{DropH}, (h, bs))            & = (h, (h, \max_{h \leq x < h + 3} \textsc{Top}(x, bs), \texttt{H}) : bs) \\
		\textsc{Eval}(\texttt{DropV}, (h, bs))            & = (h, (h, \textsc{Top}(h, bs), \texttt{V}) : bs)                         \\
		\textsc{Eval}(\texttt{MoveBefore}(E, x), (h, bs)) & = \textsc{Eval}(E, (h+x, bs))                                            \\
		\textsc{Eval}(\texttt{MoveAfter}(E, x), s)        & = (h + x, bs)\ \text{where}\ (h, bs) = \textsc{Eval}(E, s)               \\
		\textsc{Eval}(E; E', s)                           & = \textsc{Eval}(E', \textsc{Eval}(E, s))                                 \\
		\textsc{Eval}(\texttt{Loop}(1, E), s)             & = \textsc{Eval}(E, s)                                                    \\
		\textsc{Eval}(\texttt{Loop}(x, E), s)             & = \textsc{Eval}(\texttt{Loop}(x-1, E), \textsc{Eval}(E, s))              \\
		\textsc{Eval}(\texttt{Embed}(E), (h, bs))         & = (h, bs')\ \text{where}\ (h', bs') = \textsc{Eval}(E, s)                \\
		\textsc{Top}(bs, x)                               & = \max \{ y ~|~ (b, x', y) \in bs, x=x'\}
	\end{align*}
	\caption{Syntax and semantics of the tower building DSL.}
	\label{fig:tower-syntax-semantics}
\end{figure}

\subsubsection{Synthesis Problem}
The input to the synthesizer is a set of blocks $B$ where each block is represented as a tuple $(b, x, y)$. Here, $b \in \{\texttt{H}, \texttt{V}\}$ is denotes the type of the block (either horizontal 1x3 or vertical 3x1) and $(x, y)$ is the block's position. $B$ must be a valid tower, which means that the blocks must not overlap.
Given this input,
the synthesis problem is to produce a program $P$ such that $\textsc{Eval}(P, s_0) = B,$ where $s_0 = (0, [~])$ is the initial state with no blocks placed and the hand at $x = 0$.

\subsubsection{Distance Function}\label{sec:tower-dist}

The distance function for the tower building domain is based on the insight that translating a tower along the x-axis is straightforward, so we want our distance function to be \emph{translation-invariant}. Hence, two programs that build the same tower in nearby places should be deemed similar. Based on this intuition, we use the Jaccard distance to compare two towers, and we normalize them before we compare them so that their leftmost block is at $x=0$. Specifically, the distance metric between states is defined as follows: \[\delta(s, s') = \delta_J(z(s), z(s')),\] where $\delta_J$ is the Jaccard distance and $z$ is the normalizing function defined as:
\[z((h, bs)) = (h, \{ (b, x-x_{min}, y) ~|~ (b, x, y) \in bs \}.\]

\subsubsection{Rewrite Rules}
Similar to the previous domains, some of the constructs in the tower building DSL take integer valued arguments. Hence, we use rewrite rules that allow incrementing and decrementing these integers, as in the inverse CSG and regular expression domains. These rules suffice to change loop iteration counts and to modify the movement operators.
\section{Implementation}\label{sec:implementation}

We have implemented our proposed synthesis technique in a new tool called \name written in OCaml. In what follows, we describe some optimizations over the basic synthesis algorithm presented in Section~\ref{sec:algorithm}.

\paragraph{Randomization} Our implementation of the {\sc MetricSynth} algorithm is randomized and calls the {\sc ExtractTerm} and {\sc Repair} procedure multiple times. In particular, \name samples multiple programs accepted by the XFTA by calling {\sc ExtractTerm} multiple times. Furthermore, for each extracted program, \name attempts to repair it multiple times if the {\sc Repair} procedure fails. Hence, in order to sample different programs and different repairs, we introduce randomness in both the extraction and repair procedures. Specifically, we modify the {\sc ExtractTerm} procedure to consider a randomly selected subset of the automaton transitions when generating a program accepted by the FTA.\@  Similarly, we modify {\sc Repair} to consider a random subset of the rewrites when generating candidate programs to select from. Such randomization helps compensate for the greedy nature of these algorithms by introducing the possibility of taking a locally suboptimal step that turns out to be globally optimal.

\paragraph{Incremental clustering}
Since the clustering technique is a significant cost of approximate FTA construction, our implementation performs a few modifications. In particular, instead of computing all clusters and then sorting them, it first sorts the transitions and uses the first $k$ clusters that it finds. Furthermore, because the number of transitions in $\Delta_\emph{frontier}$ can be very large in the {\sc ConstructXFTA} algorithm, our implementation incrementally collects the top states in batches. This involves evaluating the frontier multiple times, rather than storing it, but we find that, in practice, we need only a small prefix of the sorted frontier. Finally, our implementation of the \textsc{Cluster} procedure  uses an M-tree data structure~\cite{Ciaccia1997} to facilitate efficient insertion and range queries.

\paragraph{Optimizations for inverse CSG}\label{sec:inverse-csg-opt}

Our instantiation of \name in the inverse CSG domain incorporates three low-level optimizations. First, it represents images as packed bitvectors to reduce their size. Second, our evaluation function for the  CSG DSL is memoized. Third, our implementation uses optimized (and, where possible, vectorized) C implementations for  bitvector operations, distance functions, and for CSG operators such as \texttt{Repeat}.

\paragraph{Optimizations for regular expressions.}\label{sec:regex-opt}

In our implementation of the regular expression domain we view the match sets as graphs where the positions in the string are the nodes and the matches are the edges. We represent these graphs as adjacency matrices using an efficient packed Boolean representation. This representation is particularly effective for synthesizing regular expressions from examples, because the example strings tend to be short, which mitigates the $O(n^2)$ memory cost of the matrix.
We also note that $\texttt{Repeat}(E, n)$ matches $(i, j)$ if $E$ matches $(i, k_1), (k_1, k_2), \dots, (k_{n-1}, j)$. That is, $\texttt{Repeat}(E, n)$ matches $(i, j)$ if there is a walk of length $n$ in the match graph for $E$ from $i$ to $j$. The walks of length $n$ are given by the $n$th power of the adjacency matrix $A_E^n$. Therefore, we can build efficient implementations for the \texttt{Repeat*} operators and for \texttt{Concat} using an efficient Boolean matrix multiplication primitive.


\section{Evaluation}

In this section, we describe a series of experiments to empirically evaluate our approach. In particular, our experiments are designed to evaluate the following key research questions:

\begin{enumerate}[leftmargin=*]
	\item {\bf RQ1:} How does \name compare against other domain-agnostic and domain-specific synthesis tools in the inverse CSG, regular expression, and tower building domains?
	\item {\bf RQ2:} How much does similarity-based clustering help with search space reduction?
	\item {\bf RQ3:} What is the relative importance of the various ideas comprising our approach?
	\item {\bf RQ4:} How do different components of our synthesis algorithm contribute to running time?
\end{enumerate}

\begin{figure}
	\centering
	\[\mathcal{U} = \{ v[x, y] = b ~|~ v \in \textsf{Image}, x, y \in \mathbb{N}, b \in \mathbb{B} \} \cup \{ v = c ~|~ c \in \textsf{Type}(v) \} \cup \{true, false\}\]
	\begin{align*}
		\aeval{f(v_1 = c_1, \dots, v_n = c_n)}                            & = (v = \eval{f(c_1, \dots, c_n)}) \\
		\aeval{(v[x, y] = true) \cup p} = \aeval{p \cup (v[x, y] = true)} & = (v[x, y] = true)                \\
		\aeval{(v[x, y] = false) - p} = \aeval{p - (v[x, y] = true)}      & = (v[x, y] = false)               
	\end{align*}
	\caption{Abstract semantics for the inverse CSG instantiation of \textsc{AFTA}. Predicates come from the universe $\mathcal{U}$.}\label{fig:abs-xformers}
\end{figure}
\begin{figure}
	\centering
	\begin{align*}
		\mathcal{U} & = \{ longestmatch(m) \leq k ~|~ m \in \textsf{Match}, k \in \mathbb{N} \} \cup \{ k \leq firstmatchpos(m) ~|~ m \in \textsf{Match}, k \in \mathbb{N} \} \\
		            & \cup \{ v = c ~|~ c \in \textsf{Type}(v) \} \cup \{true, false\}
	\end{align*}
	\begin{align*}
		\aeval{f(v_1 = c_1, \dots, v_n = c_n}                              & = (v = \eval{f(c_1, \dots, c_n)})     \\
		\aeval{\texttt{Optional}(true)}                                    & = (0 \leq firstmatchpos(m))           \\
		\aeval{(longestmatch(m) \leq k) \land (longestmatch(m) \leq k')}   & = (longestmatch(m) \leq \min(k, k'))  \\
		\aeval{(k \leq firstmatchpos(m)) \land (k' \leq firstmatchpos(m))} & = (\max(k, k') \leq firstmatchpos(m)) \\
		\aeval{\texttt{RepeatRange}((longestmatch(m) \leq k), true, n)}    & = (longestmatch(m) \leq k \times n)
	\end{align*}
	\caption{Selected abstract semantics for the regular expression instantiation of \textsc{AFTA}. Predicates are drawn from the universe $\mathcal{U}$.}\label{fig:abs-regex}
\end{figure}

\subsection{Inverse CSG}

To perform our evaluation in the inverse CSG domain, we compare \name against the following three baselines:
\begin{itemize}[leftmargin=*]
	\item {\sc Sketch-Du}: Since prior work on inverse CSG is based on the {\sc Sketch} synthesis system, we implement a baseline that uses an encoding similar to the one used in InverseCSG~\cite{Du2018}. However, since that work focuses on 3D shapes, we modify the encoding to work with our DSL for 2D geometry and also add  support for repetition (see Section~\ref{sec:related}). In addition, unlike the encoding in ~\cite{Du2018}, our sketches do not contain hints about the location of primitive shapes, so the synthesizer needs to discover the numeric parameters of these primitive shapes.
	\item {\sc FTA:} This baseline performs bottom-up synthesis (with equivalence reduction) using FTAs~\cite{dace}. Like the implementation of \name, this baseline is also implemented in OCaml. Note that this baseline only adds FTA states and transitions until a final state is reached, as our goal is to find \emph{one}, rather than all, programs consistent with the specification.
	\item {\sc AFTA:} This baseline performs bottom up synthesis with abstract FTAs (AFTAs)~\cite{blaze}. In particular, this method uses abstract values as states of the FTA, constructs  FTA transitions using the abstract semantics, and performs abstraction refinement to deal with spurious programs extracted from the AFTA. Our implementation of this  baseline is also in OCaml and uses the same matrix  abstract domain from~\cite{blaze} since bitmap images can be viewed as matrices. However, since the underlying DSLs are different, we implement the abstract transformers shown in \Cref{fig:abs-xformers} for our domain-specific language.
\end{itemize}

\paragraph{Benchmarks}  Since  prior work on Inverse CSG~\cite{Du2018,shapemod} mostly targets 3D benchmarks, we construct our own benchmark suite for 2D Inverse CSG. Specifically, we consider a total of 40 benchmarks, where 25 correspond to outputs of randomly generated programs (modulo some non-triviality constraints) and 15 are hand-written benchmarks of visual interest.

\paragraph{Setup} We run these experiments on a machine with two AMD EPYC 7302 processors (64 threads total) and 256GB of RAM.\@ We use a time limit of 1 hour and a memory limit of 4GB (so that we can run many benchmarks at the same time). For the hyper-parameters for \name, we use $\epsilon=0.2$, $w=200$, and the maximum number of rewriting steps is $n=500$.

\paragraph{Summary of results} The results of this evaluation are shown in Figure~\ref{fig:eval-main}(a).  \name is able to solve 78\% of these benchmarks, but the baselines fail on all of them except at most 3. In what follows, we discuss why the  baselines perform poorly and the failure cases for \name.

\paragraph{\textsc{FTA} results} The \textsc{FTA} baseline fails on all but the smallest of the handwritten benchmarks. When it fails, it is universally because it runs out of memory. For this domain, few programs produce exactly the same output image, so
equivalence reduction is not sufficient to reduce memory consumption.

\paragraph{\textsc{AFTA} results} We found that the \textsc{AFTA} baseline is unable to solve \emph{any} benchmarks when we include the \texttt{Repeat} operator in the DSL, as repetition causes fundamental difficulties with the abstraction refinement phase of the {\sc Syngar} technique~\cite{blaze}. However, the AFTA baseline can solve 3 of the 40 benchmarks if we omit the \texttt{Repeat} operator from the DSL. For many of the remaining benchmarks, the target  programs become quite complex without the use of the \texttt{Repeat} operator, so the AFTA approach fails either because it reaches the time limit or fails to find a program with the AST depth limit of 40.

\paragraph{\textsc{Sketch-Du} results} Our third baseline \textsc{Sketch-Du}, which is an adaptation of InverseCSG~\cite{Du2018} to our setting, is able to solve three of the handwritten benchmarks but fails on the remaining ones. For the 37 benchmarks it cannot solve, it runs out of memory 75\% of the time and runs out of time 18\% of the time. We attempted to provide this baseline with parameters that would minimize its memory use and maximize its chances of successfully completing the benchmarks. To that end, we used \textsc{Sketch}'s specialized integer solver to reduce memory overhead; we controlled the amount of unrolling in the sketch based on the size of the benchmark program, and we used \textsc{Sketch}'s example file feature, which reduces the time required to find counterexamples during the CEGIS loop. However, even with all of these optimizations, we found that the large number of examples in the inverse CSG domain (one per pixel, so 1024 total) causes \textsc{Sketch}  to perform many iterations of the CEGIS loop.

\paragraph{\name results} \name performs significantly better than the other baselines, solving all but 9 of the 40 benchmarks.
Based on our manual inspection of these 9 benchmarks, we found two dominant failure modes. One of them is that the beam width $w$ may be too narrow in some cases, causing critical subprograms to be dropped from the search space. This effect is more pronounced when the benchmark relies on subprograms that are far from the goal $O$ according to $\delta$. One pattern that we noticed among the failure cases is that they include subprograms where one shape is subtracted from another, producing a complex shape that is distant from its inputs and also distant from the final image. The second way that a benchmark can fail is that there is a program close to the solution that is contained in the XFTA, but the \textsc{Extract} and \textsc{Repair} procedures are unable to find it. While extracting all programs accepted by the XFTA could mitigate the problem, the overhead of doing so is prohibitively expensive in many cases.

\subsection{Regular Expression Synthesis}

Our second application domain  is regular expression synthesis. Given a sketch and a set of positive and negative examples, the task is to find a regular expression that conforms to the given sketch and matches all positive examples, while matching none of the negative ones.

For this application domain, we perform an empirical comparison against the following baselines:
\begin{itemize}[leftmargin=*]
	\item \textsc{Regel}: This baseline is a state-of-the-art regular expression synthesis tool~\cite{regel}. It performs \emph{top-down} (rather than bottom-up) synthesis and uses a number of SMT-based pruning strategies to reduce the search space.
	\item \textsc{FTA}: This baseline performs bottom-up enumeration with equivalence reduction using FTAs for our regular expression DSL.
	\item \textsc{AFTA}: As in the Inverse CSG domain, this baseline is an abstraction-based version of bottom-up enumeration with equivalence reduction~\cite{blaze}. We use a predicate abstraction that tracks the length of the  longest match and the beginning of the first match (see \Cref{fig:abs-regex}). These predicates effectively allow the tool to avoid examining programs which do not match the whole string or that do not match from the beginning.
\end{itemize}

We note that the FTA and AFTA baselines use the same interpreter as \name and utilize the optimization discussed in Section~\ref{sec:implementation}, which uses matrix multiplication to efficiently evaluate terms. However, {\sc Regel} uses a different regular expression matching algorithm based on the Brics automaton library, which is not as efficient as our optimized implementation for this DSL.

\paragraph{Benchmarks} For this experiment we use the Stack Overflow dataset taken from~\cite{regel}. This benchmark was collected from user questions and it contains 122 distinct tasks, each of which consists of a natural language description and a set of input-output examples. For each task, \citeauthor{regel} automatically generates a set of \emph{sketches} (i.e. partial programs) that capture additional constraints about the target regular expression that are present in the natural language description. This gives us a total of 2173 total task/sketch pairs to use in our evaluation. However, we note that some of these synthesis problems may not be solvable since the generated sketches could be wrong.

\paragraph{Experimental setup} These experiments are run on a machine with two Intel Xeon 8375C processors (with a total of 128 threads) and 256GB of RAM.\@ We use a time limit of 5 minutes and a memory limit of 4GB. We use $\epsilon=0.3$, $w=200$, and $n=100$ for the regular expression domain.

\paragraph{Results summary} The results of this experiment are summarized in Figures~\ref{fig:eval-main}(c) and ~\ref{fig:eval-main}(d), where the latter figure ``zooms in" on the harder benchmarks.  Among the four tools, \name achieves the best performance, solving 76\% of the regex benchmarks, compared to 61\% of {\sc Regel}. The FTA baseline significantly outperforms the abstraction-based approach, solving 66\%  (for FTA) compared to 13\% (for AFTA). One caveat about these results is that, while the comparison between \name, FTA, and AFTA is apples-to-apples,   {\sc Regel} uses  a different (and less efficient) interpreter for their  DSL.~\footnote{We believe it is due to this implementation difference in the interpreter that FTA slightly outperforms {\sc Regel}. The pruning heuristics used in {\sc Regel} allow it to scale without needing a custom interpreter.} Another caveat is that, while our predicate abstraction does not yield good results, it \emph{may} be possible to construct abstractions that perform better in the regex domain. Nevertheless, we believe these results substantiate our claim that  (a) \name is competitive with  state-of-the-art tools for the regex domain, and (b) metric program synthesis yields improvements over basic observational equivalence reduction.

\subsection{Solving Tower Building Tasks}
As our third application domain, we consider tower building tasks that are inspired by planning in AI and that were used for evaluating program synthesis tools in prior work~\cite{blended,dreamcoder}. As explained in Section~\ref{sec:inst}, the goal of this task is to generate a program that constructs a given configuration of blocks.

For this domain, we compare \name against two baselines:

\begin{itemize}[leftmargin=*]
	\item {\sc Neural:} Our first baseline is a state-of-the-art neural synthesizer~\cite{blended} that combines top-down program synthesis with a neural network that predicts the possible outputs of a partial program. Since this baseline is trained on a set of representative tower building tasks, it can utilize tower motifs encountered during training to solve new tasks.
	\item {\sc FTA:} As in the previous two domains, our second baseline performs bottom-up enumerative search with equivalence reduction using FTAs.
\end{itemize}

For this domain, we do not compare \name against the abstraction-based FTA approach, as the combination of loops and mutable state make this domain difficult grounds for applying prior work~\cite{blaze}. In fact, we believe that applying abstraction refinement techniques to this domain/DSL is an open research problem in its own right.

\paragraph{Benchmarks}
Our tower building benchmarks are drawn from~\cite{blended}, which are constructed by systematically composing tower building programs together (e.g., taking a program that builds a $w\times h$ bridge and building two side-by-side, varying the size and spacing). The original benchmark suite contains 40 tasks, but we found that two of the tasks are duplicates and four are not expressible in the DSL described in~\cite{blended}, as they require loops with non-constant iteration counts. We remove these six tasks, resulting in a total of  34 tower building benchmarks used in our evaluation.

\paragraph{Experimental setup}  These experiments are performed on a machine with two Intel Xeon 8375C processors (128 threads total) and 256GB of RAM.\@ We use a time limit of 10 minutes and a memory limit of 4GB. For hyper-parameters, we use $\epsilon=0.4$, $w=100$, and $n=100$.

\begin{figure}
    \centering
    \begin{subfigure}{.5\textwidth}
  \centering
  \includegraphics[width=\linewidth]{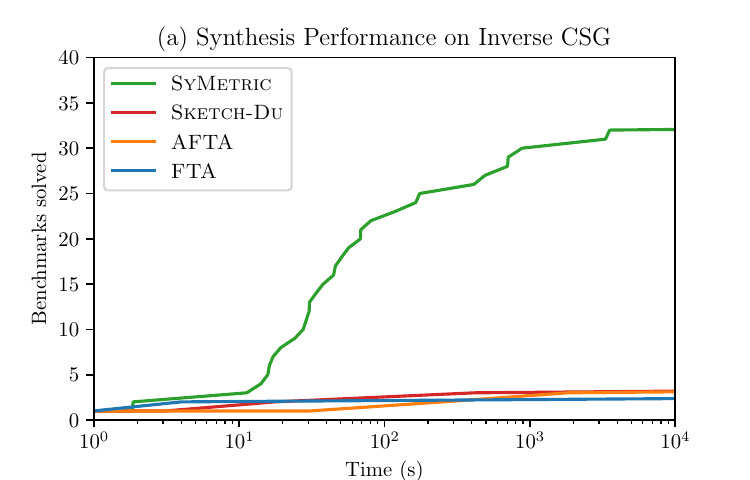}
\end{subfigure}%
\begin{subfigure}{.5\textwidth}
  \centering
  \includegraphics[width=\linewidth]{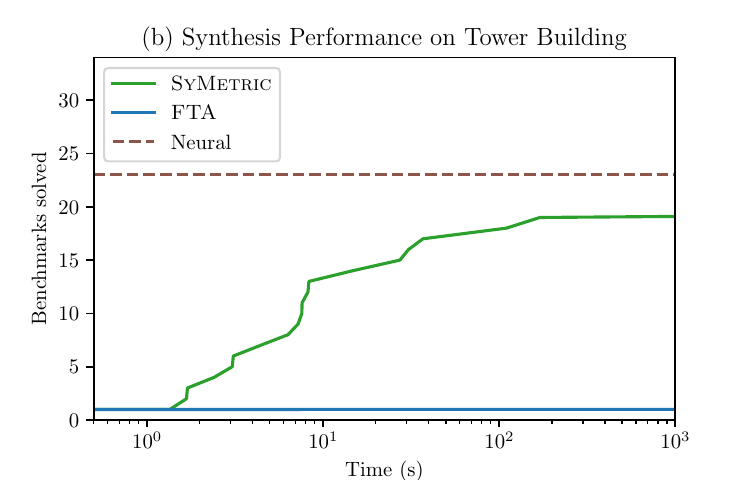}
\end{subfigure}
\begin{subfigure}{.5\textwidth}
  \centering
  \includegraphics[width=\linewidth]{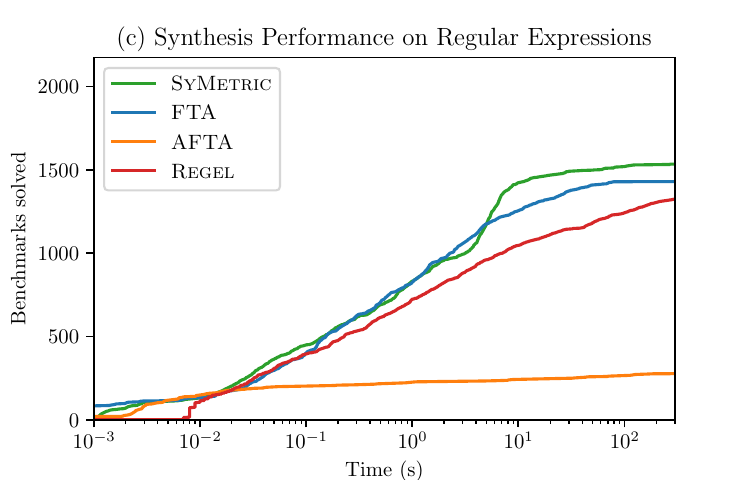}
\end{subfigure}%
\begin{subfigure}{.5\textwidth}
  \centering
  \includegraphics[width=\linewidth]{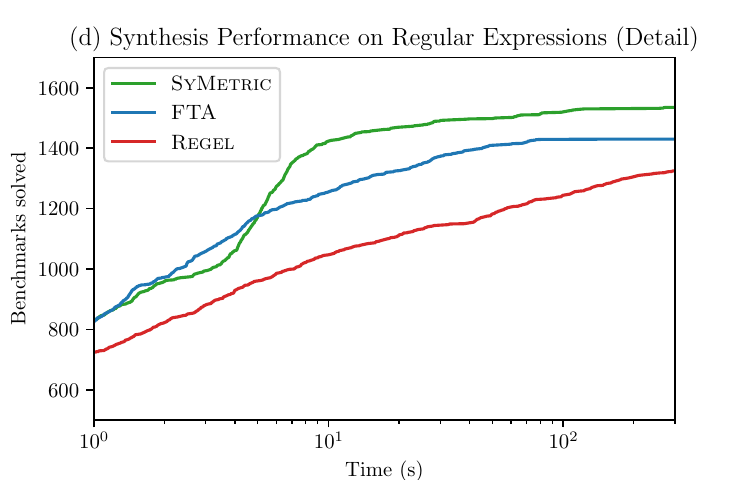}
\end{subfigure}
\caption{Synthesis performance on the inverse CSG, tower building, and regular expression domains.}\label{fig:eval-main}
\end{figure}

\paragraph{Results summary} The results for this domain are summarized in Figure~\ref{fig:eval-main}(b). Note that we do not show running time for \textsc{Neural}, as it is not reported in~\cite{blended} and we do not have access to their model. Overall, we find that \name approaches the performance of the neural approach and that it performs significantly better than \textsc{FTA}. In particular, \textsc{FTA}  performs poorly in this domain because the target programs tend to be fairly large and not many of the enumerated programs result in the same block configuration, so equivalence reduction is not as effective in this context. On the other hand, metric program synthesis can deal with the large size of the search space by exploiting observational similarity and by using ranking to select promising subprograms. Finally, we note that, while \textsc{Neural} is able to solve a few more benchmarks compared to \name, it can do so by using  motifs learned from training data as building blocks. In contrast, \name can achieve similar performance without requiring access to large amounts of training data.

\subsection{Detailed Evaluation of Metric Synthesis}

To gain more insights about the effectiveness of metric program synthesis and answer our research questions RQ2-RQ4, we perform a more detailed evaluation of \name in the inverse CSG domain. In particular, we explore distance-based clustering in more depth and present the results of relevant ablation studies.

\subsubsection{Effectiveness of Clustering}\label{sec:clustering-eval}

\begin{figure}
	\includegraphics[width=0.8\textwidth]{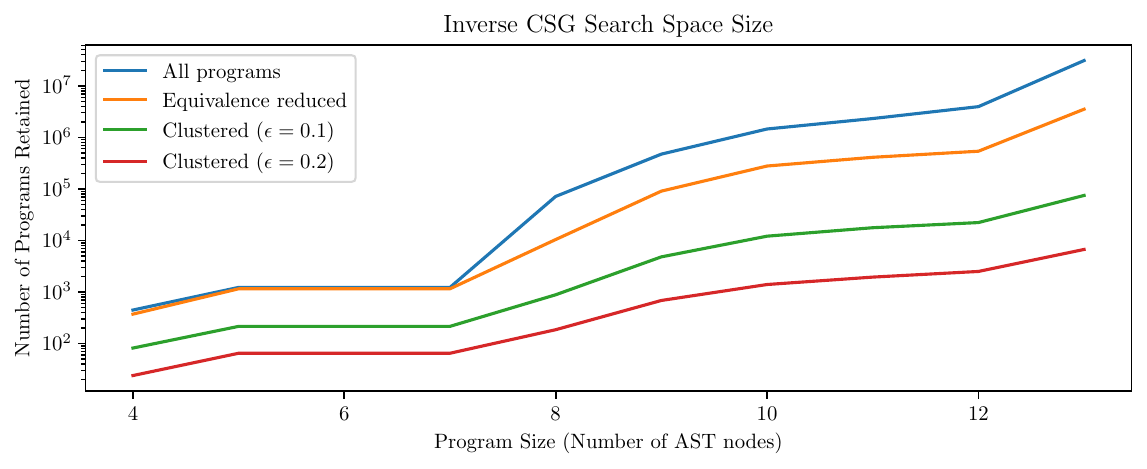}
	\caption{Size of the CSG program space for programs with up to $n$ AST nodes.}\label{fig:search-space}
\end{figure}

To evaluate the benefits of similarity-based clustering, we perform the following experiment on the inverse CSG domain:

\begin{enumerate}[leftmargin=*]
	\item First, we generate a set of programs with $n$ non-terminals.
	\item Then, for each value of $n$, we apply equivalence reduction to remove equivalent programs. \item Finally, we group the set of distinct programs  using \Cref{alg:clustering}.
\end{enumerate}
\Cref{fig:search-space} shows the number of clusters for two different values of $\epsilon$, namely $\epsilon=0.1$ and $\epsilon=0.2$. We only consider programs with up to $n=13$ AST nodes, because enumerating all programs for larger values of $n$ is not computationally feasible.

As is evident from  \Cref{fig:search-space}, equivalence reduction reduces the number of programs that must be retained by approximately one order of magnitude, and similarity-based clustering reduces the search space even more dramatically.   In particular, for $\epsilon=0.1$, there is an approximately $10\times$ reduction compared to just grouping based on equivalence and an even larger reduction for the coarser $\epsilon$ value of $0.2$. Hence, this experiment shows that there are many programs that are observationally similar but not equivalent, which partly explains why metric program synthesis is effective in this domain.

\subsubsection{Ablation Studies}

\begin{figure}
	\centering
	\includegraphics[width=0.8\textwidth]{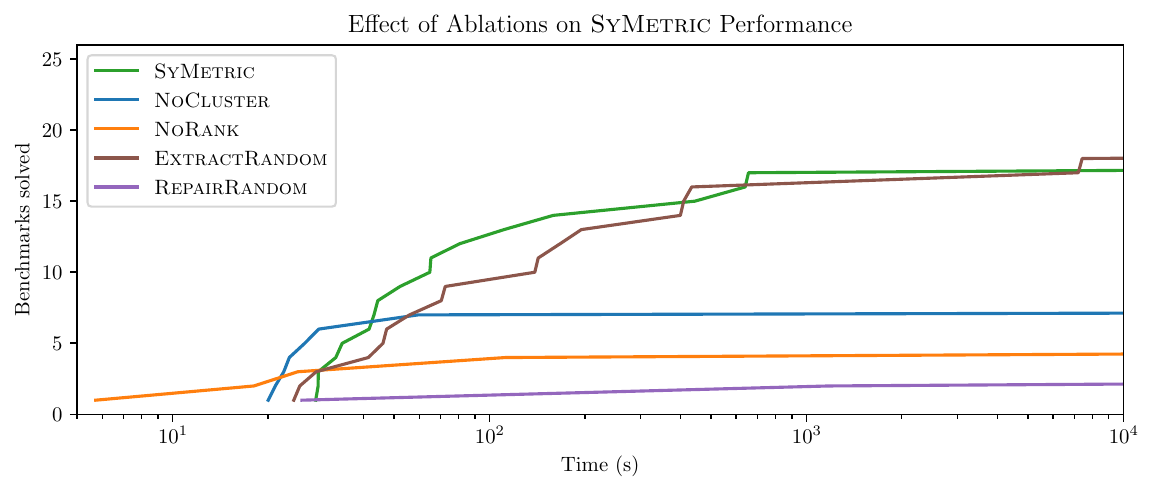}
	\caption{Effect of our ablations on \name for inverse CSG.}\label{fig:ablations}
\end{figure}

In this section, we describe a set of ablation studies to evaluate the relative importance of different algorithms used in our approach. We consider the following ablations:

\begin{itemize}[leftmargin=*]
	\item {\sc NoCluster}: This is a variant of \name that does not perform clustering during FTA construction. However, it still performs repair after extracting a program from the FTA.
	\item {\sc NoRank}: This is a variant of \name that does not use distance-based ranking during XFTA construction. Instead, it picks $w$ randomly chosen (clustered) states to add to the automaton in each iteration rather than ranking them according to the distance metric and picking the top $w$.
	\item {\sc ExtractRandom}: This variant does not use our proposed distance-based program extraction technique. Instead, it randomly picks programs that are accepted by the automaton. (However, note that the final states of the XFTA are still determined using the distance metric.)
	\item {\sc RepairRandom}: This variant does not use our distance-based program repair technique. Instead, after applying a rewrite rule during the {\sc Repair} procedure, it randomly picks one of the programs rather than using the distance metric to pick the one closest to the goal.
\end{itemize}

The results of these ablation studies are presented in~\Cref{fig:ablations} (again, for the inverse CSG domain). We find that, for this domain, the most important component of our algorithm is the distance-guided \textsc{Repair} procedure, followed by ranking during XFTA construction, and the use of clustering. The distance-guided \textsc{Extract} procedure seems to have less impact, but there is still a noticeable increase in synthesis runtime  for shorter running benchmarks if we randomly choose a program instead of using the distance metric for extraction.

Disabling ranking and clustering both yield noticeable performance improvements for some of the easier benchmarks. This is because both ranking and clustering are relatively expensive parts of the synthesis algorithm ({see Section~\ref{sec:eval-time}.}). While ranking is cheap on its own, if it is disabled, we only need to enumerate new states until we can build $w$ clusters. In contrast, when ranking is enabled, we need to look at the entire frontier at least once so we can sort it. Similarly, disabling clustering is a significant time saver for easier benchmarks. However, both clustering and ranking have a huge positive impact for the harder benchmarks. In fact, without them, the number of benchmarks solved within the 1 hour time limit drops very significantly.

\subsubsection{Detailed Evaluation of Running Time}\label{sec:eval-time}
In this section, we explore the impact of different sub-procedures on running time, again on the inverse CSG domain. Specifically, Figure~\ref{fig:method-times} compares the running times of XFTA construction (the {\sc ConstructXFTA} procedure), program extraction ({\sc Extract}), and program repair (procedure {\sc Repair}). Since our synthesis algorithm calls {\sc Extract} and {\sc Repair} multiple times, we show the aggregate running time of these procedures across all calls.

\begin{figure}
	
\begin{tabular}{lrrrrrr}
\toprule
Benchmark & \multicolumn{2}{c}{\textsc{ConstructXFTA}} & \multicolumn{2}{c}{\textsc{Extract}} & \multicolumn{2}{c}{\textsc{Repair}} \\
& Median & Max & Median & Max & Median & Max \\
\midrule

Generated & 35.5 & 40.9 & 0.1 & 12.1 & 21.2 & 664.0 \\
Hand-written & 15.4 & 65.1 & 0.0 & 21.2 & 6.0 & 3481.6 \\
All & 29.5 & 65.1 & 0.1 & 21.2 & 10.2 & 3481.6 \\
\bottomrule
\end{tabular}

	\caption{Runtime breakdown (in seconds) for  different sub-procedures of \name on inverse CSG. }\label{fig:method-times}
\end{figure}

In most cases, the running time of the {\sc ConstructXFTA} procedure dominates total synthesis time.  In contrast, program extraction from the XFTA using our greedy approach is quite fast, taking a median of about $0.1$ seconds. Finally, while the average  running time of \textsc{Repair} is around 10 seconds, it varies widely depending on how many calls to {\sc Repair} are made and how many rewrite rules we need to apply to find the correct program.

\begin{figure}
	
\begin{tabular}{lrrrrrr}
\toprule
Benchmark & \multicolumn{2}{c}{Expansion} & \multicolumn{2}{c}{Clustering} & \multicolumn{2}{c}{Ranking} \\
& Median & Max & Median & Max & Median & Max \\
\midrule

Generated & 21.9 & 25.4 & 3.9 & 8.1 & 0.0 & 0.1 \\
Hand-written & 7.9 & 19.9 & 4.3 & 36.5 & 0.0 & 0.2 \\
All & 18.5 & 25.4 & 4.0 & 36.5 & 0.0 & 0.2 \\
\bottomrule
\end{tabular}

	\caption{Runtime breakdown (in seconds) for  different sub-procedures of \textsc{ConstructXFTA} on inverse CSG. }\label{fig:eval-xfta-times}
\end{figure}

Figure~\ref{fig:eval-xfta-times} provides a more detailed look at XFTA construction. Recall that {\sc ConstructXFTA} consists of three phases, namely expansion, clustering, and ranking. In Figure~\ref{fig:eval-xfta-times}, we show the running time of each of these phases during XFTA construction. As we can see in this table, the expansion phase dominates XFTA construction time. This is not surprising because expansion requires evaluating DSL programs to construct new states. Ranking is extremely fast and barely takes any time. Clustering takes around four seconds on average, although there are some outlier benchmarks where clustering ends up being more expensive than the expansion phase.

\section{Related Work}\label{sec:related}

Our work is related to and builds on several different lines of work that we discuss here.

\paragraph{Bottom-up Synthesis}
Our work builds heavily on bottom-up synthesis with equivalence reduction, an idea that  was introduced concurrently by~\citeauthor{AlbarghouthiGK13} and~\citeauthor{transit}. Later work by ~\citeauthor{dace} explored another variation of this idea in the context of version space learning and showed how to use Finite Tree Automata (FTA) to compactly represent the space of programs consistent with a given specification. Our work was particularly inspired by \textsc{Blaze}~\cite{blaze} which demonstrated the use of abstraction refinement to speed up bottom up search. Abstractions provide a mechanism for grouping closely related solutions, allowing large sets of solutions to be ruled out by evaluating only one abstract solution. In this regard, \textsc{Blaze} can be viewed as performing equivalence reduction over abstract domains. In this work, we explore another relaxation of observational equivalence based on the notion of observational similarity rather than abstract equivalence. We believe that our metric program synthesis idea is complementary to the abstraction refinement approach and can work well in settings for which abstract domains are hard to design or where abstractions do not effectively reduce the search~space.

\paragraph{Quantitative Synthesis}
A key part of our algorithm is the use of distance metrics  to group similar programs  and to rank them. In this regard, our method bears similarities to prior work  exploring the use of quantitative goals in program synthesis, both in the reactive synthesis space~\cite{CernyH11} and in  functional synthesis (e.g. ~\cite{Schkufza2013,Schkufza0A14}). However, in many of these cases, the quantitative objectives are used to deal with noisy or probabilistic specifications~\cite{raychev-2016-learn-progr-noisy-data,handa-2020-induc-progr-synth-over-noisy-data}, whereas we use them to perform search more effectively.

\paragraph{Neural-guided Synthesis} More recently, there has been significant interest in neural-guided program synthesis, where a neural network is trained to guide the search for a program that satisfies a specification. In early incarnations of this idea, the neural network was used simply to select components that were likely to be used by the program~\cite{Balog2017}, but starting with the work of \citeauthor{Devlin2017}, the neural network has been heavily involved in directing the search. Especially relevant to our work is the work on execution guided synthesis~\cite{ChenLS19,Ellis2019}, which uses the state of the partially constructed program to determine the most promising next step for the synthesizer. The work by \citeauthor{Ellis2019} in particular inspired the ranking phase of our current algorithm. That work uses a learned value function which is trained to evaluate the output of many different candidate programs to determine which ones to keep as part of the beam.
A common limitation of all the neural guided synthesis approaches is that they require significant work ahead of time to collect a dataset and train the algorithm on that dataset. In contrast, our approach relies on a domain-specific distance function, and we find that simple distance functions often work fairly well. There is a potential for future work that seeks to apply the insights of this work in a deep learning context.

\paragraph{Diversity}
One effect of the grouping performed by our algorithm during search is to increase the diversity of the programs in the beam, allowing it to cover more distinct programs and increasing the chance that a program close to the correct program will be included in the beam. There has been some prior work on the use of diversity measures as part of a search procedure. For example, the genetic programming community has long recognized that diversity in a population of programs is crucial to avoid converging to low-quality local minima and has explored a number of diversity measures to maintain diversity of the population~\cite{GenDiversity}. Similarly, in the context of beam search for NLP, there has been recent recognition that the top-K elements of a beam may be too close to each other and fail to capture multiple modes in the underlying distribution, which has led to the proposal of \emph{Diverse Beam Search} to force proposals in a beam to be sufficiently different from each other~\cite{DBS_2018}. Our work shares some intuitions with some of these prior works, but to our knowledge, this paper is the first to apply the idea of grouping based on a similarity function in the context of FTA-based synthesis.

\paragraph{Program Synthesis for Inverse CSG}
There has been a lot of interest in the CAD community in using program synthesis techniques to reverse engineer CAD problems. Two early works in this space are InverseCSG~\cite{Du2018} and the work of \citeauthor{NandiWPBGT18}.
Both aim to reverse engineer 3D CSG programs from meshes, but both rely on specialized algorithms to do a significant part of the work. For example, \citeauthor{NandiWPBGT18} rely on a set of domain specific oracles that examine the mesh and generate proposed decompositions (i.e. splitting the mesh into a union of two simpler shapes). These oracles are powerful but highly specialized to the CSG domain. Similarly, InverseCSG relies on a specialized preprocessing phase to identify all constituent primitive shapes and their parameters, so the synthesizer only has to discover the Boolean structure of the shape. In contrast, our synthesis method can solve for all primitives and their parameters without relying on a domain-specific preprocessor.
Additionally, InverseCSG relies on a segmentation algorithm to break a large shape into small fragments that are then assembled into the final shape. In contrast, we aim to solve the entire inverse CSG problem as a single synthesis task. Finally, our program space is richer than that of InverseCSG because it includes a looping construct in addition to the primitives and Boolean operations. Our experimental results show that we can do with a single algorithm what prior work required an entire pipeline of complex and very specialized algorithms.

In addition to the program synthesis oriented work on inverse CSG, there is a line of work that focuses on the use of neural networks to recover structured models from unstructured input~\cite{csgnet,shape2prog}. Their performance is generally very good but dependent on  training data (e.g., Shape2Prog~\cite{shape2prog}) is specialized to just furniture shapes). One of the goals of our algorithm is to only require  limited domain knowledge, which are captured through  the distance function and  repair rules. However, we believe that our approach can complement the neural methods, either by using a learned distance metric or by using a network to predict a likely space of programs, as in~\cite{Lee2018}.

There is also prior work that processes CSG programs, either to extract common structure~\cite{shapemod} or to capture regularity~\cite{NandiWAWDGT20}. ShapeMOD~\cite{shapemod} is a technique for extracting common macros from a library of CSG programs. These macros form a domain specific language for a particular class of CSG programs (e.g., furniture) and help provide more physically plausible results that are biased towards patterns common in the target domain. Our method does not rely on a set of training programs but could utilize macros (like the ones learned by ShapeMOD) if they were available. \citeauthor{NandiWAWDGT20} show that it is possible to post-process the output of an InverseCSG-like system in order to extract loops, which produces programs that are more general and easier to modify. However, such an approach first requires  synthesizing the loop free program, which can get quite large for models with a lot of repetition.

\paragraph{Synthesis of Regular Expressions} There is a long history of work on synthesizing regular expressions from examples. Recent work includes \textsc{Regel}~\cite{regel} and \textsc{AlphaRegex}~\cite{alpha-regex}, both of which are top-down synthesizers that employ upper and lower bounds for pruning. \textsc{Regel} additionally leverages natural language descriptions to improve generalizability. In particular, \textsc{Regel} parses the natural language descriptions into so-called \emph{hierarchical sketches} and uses top-down enumerative search combined with  SMT-based pruning to find a sketch completion that satisfies the examples. In our evaluation, we utilize the sketches generated by {\sc Regel} but solve those sketches using metric program synthesis instead of top-down SMT-guided search.

Repair of regular expressions, which is related to our local search, has also attracted significant attention from the research community~\cite{rfixer,rebele-2018-addin-missin-words-to-regul-expres}. Because our method starts the local search process with programs that are already fairly close to the goal, our method can use simpler rewrite-based techniques for repair.

\paragraph{Synthesis of Tower Building Programs}
The tower building domain first appears in~\cite{dreamcoder} and is used as a benchmark in~\cite{blended}. The appeal of this domain comes from its use of  loops and mutable state, as well as its connection to classical AI planning tasks. Prior work has focused on the application of neural guided synthesis to this domain, whereas we show that a non-neural approach can also perform quite well.


\section{Conclusion}\label{sec:conclusion}

We  presented a new synthesis technique, called \emph{metric program synthesis}, that performs search space reduction using a distance metric. The key idea behind our technique is to  cluster similar states during bottom-up enumeration and then perform program repair once a program that is ``close enough'' to the goal is found. In more detail, our approach constructs a so-called \emph{approximate finite tree automaton} that  represents a set of programs that ``approximately'' satisfy the specification. Our method then repeatedly extracts  programs from this set and uses distance-guided rewriting to find a repair that exactly satisfies the given input-output examples.

Our proposed synthesis algorithm is intended for domains that have two key properties: (1) the DSL contains many programs that are semantically similar, and (2) programs that are semantically similar also tend to be syntactically close. With this intuition in mind, we have  instantiated our synthesis framework in three different domains (inverse CSG, regular expression synthesis, and tower building) by defining suitable distance metrics. Our evaluation shows that our tool, \name, outperforms prior domain-agnostic FTA-based techniques in these domains. Furthermore, we compare our approach against domain-specific synthesizers and show that the performance of \name is competitive with these tools despite not utilizing any training data or domain-specific synthesis algorithms.


\clearpage
\bibliography{localrefs,references}


\end{document}